\documentclass[screen,manuscript,acmsmall,nonacm]{acmart}
\usepackage{verbatim}

\usepackage{enumitem}

\usepackage{multirow}

\usepackage{graphicx}
\usepackage{textcomp}
\graphicspath{ {./images/} }

\usepackage[utf8]{inputenc}
\usepackage{tikz}
    
\usepackage{tabularx}
\usepackage{booktabs}
\newcommand*{\nextrow}{\\\rule{0pt}{4ex}}
\newcolumntype{Y}{>{\centering\arraybackslash}X}
\newcolumntype{R}{>{\raggedleft\arraybackslash}X}
\newcolumntype{L}{>{\raggedright\arraybackslash}X}
    
\usepackage{hyperref}

\usepackage{todonotes}
\usepackage{appendix}
\usepackage{csquotes}

\AtBeginDocument{%
  \providecommand\BibTeX{{%
    \normalfont B\kern-0.5em{\scshape i\kern-0.25em b}\kern-0.8em\TeX}}}
    
\AtBeginDocument{%
  \providecommand\BibTeX{{%
    \normalfont B\kern-0.5em{\scshape i\kern-0.25em b}\kern-0.8em\TeX}}}

\setcopyright{none}

\begin{document}

\title[Critical-Reflective Human-AI Collaboration]{Critical-Reflective Human-AI Collaboration: Exploring Computational Tools for Art Historical Image Retrieval}

\author{Katrin Glinka}
\email{katrin.glinka@fu-berlin.de}
\orcid{1234-5678-9012}
\affiliation{%
  \institution{Freie Universität Berlin}
  \streetaddress{Königin-Luise-Str. 24/26}
  \city{Berlin}
  \country{Germany}
  \postcode{14195}
}

\author{Claudia Müller-Birn}
\email{clmb@inf.fu-berlin.de}
\orcid{0000-0002-5143-1770}
\affiliation{%
  \institution{Freie Universität Berlin}
  \streetaddress{Königin-Luise-Str. 24/26}
  \city{Berlin}
  \country{Germany}
  \postcode{14195}
 }

\begin{abstract}
 Just as other disciplines, the humanities explore how computational research approaches and tools can meaningfully contribute to scholarly knowledge production. Building on related work from the areas of CSCW and HCI, we approach the design of computational tools through the analytical lens of `human-AI collaboration.' Such work investigates how human competencies and computational capabilities can be effectively and meaningfully combined. However, there is no generalizable concept of what constitutes `meaningful' human-AI collaboration. In terms of genuinely human competencies, we consider criticality and reflection as guiding principles of scholarly knowledge production and as deeply embedded in the methodologies and practices of the humanities.
Although (designing for) reflection is a recurring topic in CSCW and HCI discourses, it has not been centered in work on human-AI collaboration. We posit that integrating both concepts is a viable approach to supporting `meaningful' human-AI collaboration in the humanities and other qualitative, interpretivist, and hermeneutic research areas. Our research, thus, is guided by the question of how critical reflection can be enabled in human-AI collaboration.  
We address this question with a use case that centers on computer vision (CV) tools for art historical image retrieval. Specifically, we conducted a qualitative interview study with art historians to explore a) what potentials and affordances art historians ascribe to human-AI collaboration and CV in particular, and b) in what ways art historians conceptualize critical reflection in the context of human-AI collaboration. We extended the interviews with a think-aloud software exploration. We observed and recorded our participants' interaction with a ready-to-use CV tool in a possible research scenario.  
We found that critical reflection, indeed, constitutes a core prerequisite for `meaningful' human-AI collaboration in humanities research contexts. However, we observed that critical reflection was not fully realized during interaction with the CV tool. We interpret this divergence as supporting our hypothesis that computational tools need to be intentionally designed in such a way that they actively scaffold and support critical reflection during interaction.
Based on our findings, we suggest four empirically grounded design implications for `critical-reflective human-AI collaboration': supporting reflection on the basis of transparency, foregrounding epistemic presumptions, emphasizing the situatedness of data, and strengthening interpretability through contextualized explanations.

\end{abstract}



\maketitle

\section{Introduction}
\label{sec:introduction}

In the humanities, retrieving objects and primary sources for the purpose of studying them constitutes a central ingredient of scholarly work. Art historians, for example, access objects or images of objects alongside their metadata and accompanying information. The decision-making in terms of what objects are considered `relevant' is highly dependent on the art historian's research question, field of study, and other contextualized factors. The process of scholarly image retrieval and corpus building has changed significantly with the digitization of textual and visual sources and the creation of digital repositories~(see, e.g.,~\cite{Drucker_IsThereDAH, Huistra_SelectingSources2016}). 
Art historians, just like other humanities scholars, can now query online databases for objects, sources, and literature without traveling to the collecting institution or contacting archivists or other researchers. 
%
Recently, research in digital humanities (DH) and the sub-field of digital art history (DAH) has been exploring computer vision (CV) as a mode of access to visual material (e.g.,~\cite{seguin2018replica, Ommer_SimilarityArt, Ufer_image_retrieval, VillaespesaMurphyApple, impett2020analyzing}), which is subsequently leading to the implementation of CV in ready to use tools\footnote{Examples for such tools are \href{https://www.iart.vision/}{iART}~\cite{iArt2021} or \href{https://imgs.ai/}{imgs.ai}~\cite{imgsai2023}.}.  
Such CV-based tools are designated to enable art historians to process thousands of images and to retrieve those relevant to their research question or in the context of other tasks, for example, when conducting provenance research or in order to compare images for the purpose of authorship attribution. However, the actual art-historical interpretation of the retrieved images is still considered a genuinely human competency~\cite{Klinke_2016}. 
Taking this into account, we investigate CV tools, i.e., computational tools, in the context of art history through the analytical lens of ``human-AI collaboration,'' a concept that refers to the integration of complementary competencies~\cite{Shneiderman_Maes_1997} of humans and computers. In this understanding, computational tools are not designed to produce significant and valid research results or insights by themselves. Instead, they must be paired with human intellect in order to contribute to achieving a shared goal~\cite{Terveen_1995}. 

It is considered crucial to understand how computational tools function and how they impact knowledge production to secure their `meaningful' integration in humanities research~\cite{koolen_toolcriticism_2018}. This perspective resonates with the understanding of research as a critical-reflective practice, which is equally valued in `traditional' humanities research as well as in DH and DAH (see, e.g.,~\cite{alvarado2019digital}). Even though there exist recommendations regarding `digital tool criticism' as an extension of such a critical-reflective practice in the humanities, those recommendations are mainly directed at the humanities scholars using such software (see, e.g.,~\cite{vanEs_ToolCriticism, koolen_toolcriticism_2018}). In their model for `digital tool criticism' that frames reflection as an integrative practice, Koolen et al.~\cite{koolen_toolcriticism_2018}, for example, only include one recommendation for tool development: namely, software should include documentation or an `about' page that covers aspects such as the tool's functionalities and how each of these selects, filters, and transforms data. We posit that the resulting tools tend not to actively emphasize and effectuate critical reflection \emph{during} usage, which distances the act of critical reflection from the interaction with a tool. Moreover, this distancing requires the humanities scholars to additionally access the documentation and relate the provided information to their usage. Since such documentation is usually written for a technical audience, i.e., detailing the source code, setup, and implementation, we question the effectiveness of this sort of detached technical documentation and information in terms of supporting critical-reflective human-AI collaboration. 

To summarize, the development of computational tools in the context of DH or DAH is often conceptualized as a `substitution problem'. In an existing research workflow, such as image retrieval, selected tasks are substituted by software.
%
We argue that the tendency to conceptualize tools as a substitution problem leads to prioritizing an engineering perspective, which misses out on integrating the possibilities that HCI and CSCW research provide in the context of human-AI collaboration, namely a focus on ``designing for reflection''~\cite{baumer2015reflective}. We hypothesize that tools need to be intentionally designed in such a way that they support critical reflection~\cite{Baumer:2017:human-centered-algorithm-design}. 
Based on these considerations, our work is guided by the research question: \emph{How can critical reflection be enabled in human-AI collaboration in the context of art historical image retrieval?} 


We investigate our research question through a qualitative interview study with 12 art historians. However, we also recognize that knowledge is created through use practice and might never become formalized~\cite{Fraunberger_Entanglement2020}. This makes it hard to address human-AI collaboration and associated concepts of critical reflection only through interviews. We assume that aspects of reflection might surface only during interaction with a computational tool. Therefore, we extend the interviews with a ``think-aloud''~\cite{Lewis_1982} software exploration during which the participants use a CV tool to conduct a real-world image retrieval task that is grounded in their current research. 

Through a reflexive thematic analysis (TA), we confirm that art historians are eager to interact with computational tools in a critical-reflective manner. However, the participants could not fully realize this eagerness during interaction with a CV tool. We interpret these divergences as support for our hypothesis that computational tools need to be intentionally designed in such a way that they actively support critical reflection. We conclude that computational tools should not be developed based on the assumption that humanities scholars are able to effectuate their competencies in criticality and reflection without any scaffolding, even more so when this would require extensive computational knowledge. 
Based on our findings, we derive implications for the design of `critical reflective human-AI collaboration.'

We make the following contributions with our research: We deepen the understanding of what constitutes `critical reflective human-AI collaboration' by building on and extending existing research from CSCW and HCI; we present findings from an empirical study that integrates expert interviews with art historians with observations of their interactions with a CV tool for image retrieval; and we propose four empirically grounded high-level design implications for the design of `critical reflective human-AI collaboration,' namely: 1) supporting reflection on the basis of transparency, 2) foregrounding epistemic presumptions, 3) emphasizing the situatedness of data, and 4) strengthening interpretability through contextualized explanations.

Our paper is divided into four main sections. We lay the foundation for our research by reviewing related work on human-AI collaboration, principles of reflection in HCI, and critical reflection in the humanities and DH (\autoref{sec:relatedwork}). We then explain our research method in \autoref{sec:methods}, where we first introduce our use case in the context of art history, specifically CV-based image retrieval. This is followed by a description of our study design. In \autoref{sec:findings}, we present the findings that relate to human-AI collaboration and critical reflection. Based on a discussion of our findings, we present four design implications for `critical reflective human-AI collaboration' (\autoref{sec:discussion}).

\section{Related Work}
\label{sec:relatedwork}

In the following, we provide an overview of existing research on human-AI collaboration in qualitative research. We then foreground reflection as one `meaningful' realization of human-AI collaboration. This research from CSCW and HCI provides the foundation for situating our research in the context of critical reflection in the humanities.  

\subsection{Conceptualizing Human-AI Collaboration in Qualitative Research}
\label{sec:humanAICollab}
Human-AI collaboration is not a clearly defined term in HCI and CSCW but represents more of a design concept.
Recent technological advancements in `artificial intelligence' (AI) have inspired CSCW and HCI scholars to investigate how to support `collaboration' between humans and AI-based systems (see, e.g.,~\cite{Terveen_1995, Wang_HHColltoHumanAIColl}). Terveen~\cite{Terveen_1995} provided one of the earliest conceptualizations of such collaboration; he states, \enquote{collaboration is a process in which two or more agents work together to achieve shared goals.} Terveen emphasizes the disciplinary grounding of what he calls \enquote{human-computer collaboration.} This perspective conceptually draws from both AI and HCI; thus, it combines different understandings of designing such collaboration. The design concept of collaboration has been taken up over the years and appears in various other concepts such as mixed-initiative interaction~\cite{Horvitz:1999mi}, interactive-intelligent systems~\cite{Jameson:2011fy}, or human-computer integration~\cite{Farooq_Grudin_2016}. Thus, a growing body of research deals with how humans and AI-based systems can effectively collaborate in particular application areas.
While there is a lack of research on areas of applications that directly relate to our field of study, i.e., the humanities and art history, we can build on related work on human-AI collaboration for qualitative research.

Jiang et al.~\cite{Jiang_serendipity} conducted a study on qualitative researchers' work practices and the potential benefits and challenges of AI-based tools in qualitative research. Their study confirms that the distribution of agency between humans and AI systems must be carefully balanced. They point out that the automation of qualitative analysis \textit{in its entirety} is infeasible. Instead, they suggest shifting the focus ``towards ways that AI could be a collaborator that works alongside humans rather than a delegate that performs specific tasks''~\cite{Jiang_serendipity}. Their study particularly stresses the importance of emphasizing human agency in data analysis. 
In response, Feuston and Brubaker~\cite{Feuston_Tools} suggest a more nuanced approach that considers the different stages during a qualitative analysis process as one of the defining parameters that influence whether and to what extent scholars believe that AI could support them in their research. Feuston and Brubaker also describe how the use of AI in different stages of a qualitative analysis process impacts knowledge production, namely, a shift in scalability, abstraction, and delegation~\cite{Feuston_Tools}.
Considering the different stages of qualitative analysis and carefully and purposefully conceiving human-AI collaboration tailored to a specific phase or sub-task within a research process is also echoed in a study by Chen et al.~\cite{chen_ambiguity}. Instead of building an AI-based system that predicts codes\footnote{Coding is a core aspect of qualitative analysis that refers to the process of assigning `codes' or labels to data, i.e., transcriptions of interview data or textual data collected from social media, which supports theory development. How researchers code differs and depends on the type of analysis they apply to the data. Coding is often performed by more than one researcher, sometimes in combination with an emphasis on inter-coder reliability.} for data instances, which turned out to be unfeasible, Chen et al. suggest that an AI system could help identify points of potential inconsistency and ambiguity in the assigned codes. The AI system is not expected to present the researchers with a solution or resolve the ambiguity. Instead, it will direct them to instances that need clarification or prompt them to reconsider diverging interpretations. 
Going a step further, Baumer et al. explore the application of computational approaches to the actual \textit{analysis} of texts. They similarly emphasize that the role of an AI-based system should be carefully balanced, as suggested by their approach of ``designing for interpretation''~\cite{Baumeretal:2020:Topicalizer}. Following this principle, Baumer et al. explore an alternative approach in computational text analysis by providing a visualization that supports the interpretation and reinterpretation of the data. They emphasize that algorithmic systems need to strike a balance between explicitly stating what their results mean, for example, explaining how the results come about, while intentionally leaving room for the interpretation of the results~\cite{Baumeretal:2020:Topicalizer}. In a similar approach, Hong et al.~\cite{HongFeustonBrubaker_Scholastic2022} relate their suggestion of using topic models for qualitative research to related work on topic modeling in DH. They conceptualize human-AI collaboration as a tool for assisting interpretive research~\cite{HongFeustonBrubaker_Scholastic2022}. 



In summary, 
researchers in the field of human-AI collaboration investigate how to effectively \emph{combine} human intelligence and computational capabilities, i.e., how such collaboration can \emph{augment human intellect}. Furthermore, they carefully investigate in their application contexts to what extent AI and humans can complement each other and profit from such collaboration. However, we still lack a more nuanced understanding of what `collaboration' between humans and AI means in specific contexts and what type of collaboration we want to support. We need to define what quality the collaboration should have in order to be considered `meaningful.' As one approach to achieving `meaningful' collaboration, we suggest foregrounding reflection, a solely human trait. 

\subsection{Using Reflection as Means of Human-AI Collaboration}
\label{sec:critical_reflective}


We posit that a `meaningful' realization of human-AI collaboration can be accomplished by explicitly designing AI technologies that support and encourage reflection on the users' side as a guiding principle of interaction~\cite{baumer2015reflective, BaumerEtAl_ReviewingReflection}.
Reflection is a somewhat elusive concept: the challenge lies in conceptualizing reflection epistemologically and operationalizing this abstract concept~\cite{baumer2015reflective}. Reflection can generally be defined as \enquote{reviewing a series of previous experiences, events, stories, etc., and putting them together in such a way as to come to a better understanding or to gain some sort of insight}~\cite{BaumerEtAl_ReviewingReflection}. This perspective builds upon theories of learning and work practice, where experience forms the basis for thinking about and planning for the future. In this process, we adapt and change our knowledge, behavior, and values ~\cite{Schoen_1983, Dewey1997}. 
In HCI, reflection has many manifestations that vary in scope; they range from ``revisiting'' to ``transformative'' events. The effects of such reflection include, for example, a change in behavior, understanding, or gaining insights~\cite{Slovak_ReflectivePracticum}.
Over the last few years, reflective technologies have been realized in various contexts, such as slow technology~\cite{Hallnas_Redstrom_2001}, persuasive design (e.g.,~\cite{foggBehaviorModelPersuasive2009}), and personal informatics (e.g.,~\cite{Li_Dey_Forlizzi_2010}). Hallnas and Redström introduced slow technology as \enquote{a design agenda for technology aimed at reflection and moments of mental rest rather than efficiency in performance}~\cite{Hallnas_Redstrom_2001}. They emphasize that technology can be slow in several ways, such as learning how it works, understanding why it works the way it does, using it and experiencing its behavior, or figuring out the consequences of its application~\cite{Hallnas_Redstrom_2001}. Building on that, Cox et al.~\cite{Coxetal2016} propose ``mindful interaction'' to realize a deliberate and intentional interaction. They suggest that an interactive system should contain a microboundary, i.e., a \enquote{small obstacle [$\ldots$] that prevents us rushing from one context to another.}



In line with this research, Baumer et al.~\cite{baumer2015reflective} suggest three essential characteristics that technology should provide to evoke reflection: ``breakdown,'' ``inquiry,'' and ``transformation.'' Breakdowns refer to the intentional violation of the users' expectations, which would create perplexity and, thus, lead to reflection. At the same time, incomprehension might also act as a prompt to engage in interpretation, i.e., making sense or deriving meaning from a particular output. 
The second dimension is the ``process of conscious, intentional inquiry''~\cite{baumer2015reflective}. Such inquiry can be supported by reviewing previous experiences or actions. It can be achieved by providing a designated space for inquiry, for example, in the form of a visualization that is separate from the main activities to which that inquiry relates and that enables an exchange with others to promote reflective inquiry. 
Baumer introduces the third dimension, transformation, as ``envisioning alternatives'' ~\cite{baumer2015reflective}. Reflection involves change - it is not just about examining the current state of the world or the self but also about imagining alternatives. Thus, such transformation leads to a change in how a human understands or conceptualizes the world.
Consequently, when supporting reflection in human-AI collaboration, users should be made aware of conflicting pieces of information (= breakdown) that they can interact with (= inquiry) to resolve possible conflicts or to promote insight (= transformation).
%
%


\subsection{Challenging Critical Reflection in (Digital) Humanities}
\label{sec:CV_ArtHistory}

Reflection is firmly embedded in the methodologies and practice of the humanities, often in conjunction with the notion of criticality. This includes, for example, the understanding of source criticism as a structured method that was ``spelled out in systematic guidelines for historical research and became a cornerstone of academic training''~\cite{muller2020hermeneutics_sourcecriticism} as early as the late 18th century. Similarly, reflecting on and through theory and reflecting on epistemology and methodology are central concepts in the humanities. In general terms, as Flusser put it: ``methodic reflection is a critique of science''~\cite{flusser2005thoughtreflection}.
In the following, we foreground related work that helps us illustrate how understandings of criticality and reflection as part of epistemology and knowledge production are being challenged and revisited in the context of engaging with AI, i.e., computational approaches and tools, in the humanities. 

The development of computational tools for the retrieval and remediation of objects constituted one of the earlier goals of DH\footnote{We use the term ``digital humanities'' (DH) for the sake of simplification throughout this paper. However, the efforts described here as the earlier goals of DH emerged under the term ``humanities computing'' or ``digitized humanities''. The adoption of the term ``digital humanities'' in the early 2000s was accompanied by a shift in focus: from the digitization of material and infrastructures that would serve humanities scholars primarily in terms of access to material to the understanding of DH as a ``methodological outlook''~\cite{kirschenbaum2016digital}.}~\cite{alvarado2019digital}. The results of these efforts now benefit the humanities as a whole, primarily in terms of increased access to sources and digitized material. In the context of art history, tool development has mostly been situated in the realm of what Drucker has called ``digitized art history,'' where digital tools ``are just new ways of doing old work a little faster, easier, and with greater access to more materials of all varieties''~\cite{Drucker_IsThereDAH}. In this sense, tool development is based on identifying time-consuming tasks that computers can perform more efficiently~\cite{Brey_DAH}. 

In this understanding, `digitized art history' relates to accessing objects through databases containing images, information, and data that reproduce symbolic encodings of ``expert'' knowledge~\cite{impett2020analyzing}. Such textual and ontological knowledge representations are usually grounded in long-established epistemologies and knowledge structures, i.e., ``doing old work''~\cite{Drucker_IsThereDAH}. In a critical reading of subject indexes and classification schemes, Rawson and Muñoz point out that taxonomies and guidelines for their application assume an underlying ``correct'' order that suppresses diversity of knowledge \cite{Rawson_AgainstCleaning}. Even in the digital realm, art-historical objects remain firmly situated within established epistemological frameworks (see, e.g.,~\cite{Carboni_Ontological, Doerr_2003, Dahlgren_DigitalUTurn2021}). Classification schemes for art historically relevant objects tend to emphasize the interpretation placed on the objects~\cite{enser1995progress}, which is, for example, the case with the widely used `Iconclass' system~\cite{couprie1978iconclass} that allows the classification of iconography in artworks. Consequently, keyword search in image collections largely aligns with established and canonized interpretations. As Underwood has argued in regards to search on literary texts, typing in a search term already encodes ``tacit hypotheses about the literary significance of a symbol''~\cite{Underwood_Theorizing_2014}, which also applies to text-based image retrieval. 
Such critical and reflective engagements with the hegemonic effects of information systems resonate with work from science and technology studies that point out how databases and classifications mirror and re-perform the knowledge economies of which they are a part~\cite{Waterton_STS_Archive}. Correspondingly, in the context of critical design in HCI, Feinberg et al.~\cite{Feinberg_Database_CriticalDesign} have pointed out how classification systems flatten and distort complex realities. 

Text-based retrieval has been, and still is, the predominant mode of accessing images, visual sources, and other non-textual material that is relevant to art historical research. While full-text search gives access to a textual document in itself, text-based image retrieval fully relies on knowledge representation, i.e., textual or numerical representations of an artwork or object. This is not only the case for computer-based systems: Card catalogs have long been the primary retrieval system for artworks and art-historically relevant objects in museums or archives. Such card catalogs also acted as a reference point for the design of computer-based information retrieval, for example, in terms of human-processable knowledge representation and classification~(see, e.g.,~\cite{Smith_retrieval_1981, chenhallPropositionsFutureMuseum1978})\footnote{The development of such domain-specific information retrieval systems was intensified in the 1970ies by specialized working groups in the cultural field, for example, by the Information Retrieval Group (IRGMA) of the Museums Association. Computer-based information retrieval systems were designed in line with domain-specific documentation and classification standards.}. Accordingly, art historians are well-trained in accessing artworks and other visual, non-textual sources and objects of study through such human-processable knowledge representations, both in the form of manual card catalogs and digital databases. Their competency to reflect on the underlying effects of knowledge representation and epistemic orders of knowledge in their field of study is paired with source criticism and is also applied when using digital databases for text-based image retrieval. 

By enabling access to objects in their pictorial form as digital images, CV opens up a mode of image retrieval that is not solely reliant on established and canonized text-based classifications and interpretations. While this broadens access to image collections and might circumvent canonized classifications and orders of knowledge, it is important to acknowledge that CV inserts new and potentially unnoticed distortions that need to be understood, accounted for, and made explicit. Despite this caveat, the potential of CV for art historical research is evident: Instead of being confined to a keyword search based on the symbolic encodings of `expert' knowledge \emph{about} the objects, art historians can search image collections using an image as input. 
This way, CV might also reduce barriers associated with cross-language search, which is particularly promising in light of a growing interest in non-western art history~\cite{zorich2013digital}. CV also holds the potential to enable exploratory approaches to image retrieval. This is helpful in art historical scholarship since researchers do not always know what they are looking for as well as they think~\cite{Underwood_Theorizing_2014}. 

In acknowledging that CV introduces new layers of symbolic encodings and knowledge economies, it also becomes obvious that it is crucial to scrutinize its distorting and hegemonic effects in order to secure `meaningful' integration in art historical research and practice. Hence, researchers need to understand and reflect on how this technology affects epistemic assumptions, research processes, and knowledge production and how the technology shapes their ``ways of thinking''~\cite{Ruhleder1995}. 
This requirement also extends to the ways in which the images and object data are mediated through interfaces and visual arrangements. As Drucker put it: ``an interface \emph{is} information, not merely a means of access to it''~\cite{drucker2014graphesis}. In this context, the need to critically reflect on the implications of interacting with technology is evident: using computational tools includes ``[reflecting] on the methods and premises that shape our approach to knowledge and our understanding of how interpretation is framed. Digital humanities projects are [$\ldots$] occasions for critical self-consciousness''\cite{drucker2004speculative}. 

The works reviewed above highlight how engagement with AI and computational tools in the (digital) humanities challenges critical reflection. However, what \emph{exactly} constitutes instances or practices of criticality and reflection and how they can be operationalized in practices of use is rarely made explicit~\cite{Fisher_demystifying_critical}. 
Concrete approaches to applying critical reflection in DH are emerging in the context of `tool criticism.' Here, critical reflection is framed as a means to understand \enquote{that research questions, methods, tools, and data are interdependent and choices regarding them are shaped in an interactive and reflective research process}~\cite{koolen_toolcriticism_2018}. Tool criticism, in this sense, refers to a technical reflection of the tool and the critique of the research output while also considering the tool's influence on the research
process~\cite{vanEs_ToolCriticism}. This stance echoes work that emphasizes the importance of critically thinking \enquote{about how machine learning is being designed and deployed in the specific problem domains represented by the informating, augmenting and automating of digital humanities}~\cite{Bassett2017CriticalDH}. 
However, \textit{how} such acknowledgments of the importance of criticality and reflection could be translated into actually designing AI technologies that ``scaffold the reflection process''~\cite{Slovak_ReflectivePracticum} is still under-investigated. 
\section{Research Method}
\label{sec:methods}

Our work contributes to a deepened understanding of this challenge, i.e., we investigate \emph{how critical reflection can be enabled in human-AI collaboration}. We explore this topic with a use case on CV-based image retrieval in art history and conduct a qualitative interview study. However, we also recognize that knowledge is created through use practice and might never become formalized~\cite{Fraunberger_Entanglement2020}. Consequently, we consider it insufficient to address human-AI collaboration and associated concepts of critical reflection only through interviews. Therefore, we extend the interviews with a think-aloud software exploration during which the participants use a CV-based tool to conduct a real-world image retrieval task that is grounded in their current research.

This method setup allows us to investigate our overarching research question along four sub-questions:
\begin{enumerate}
    \item What potential and affordances do art historians ascribe to human-AI collaboration (specifically regarding CV)?
    \item How do these potentials and affordances surface when art historians interact with a CV-based image retrieval tool?
    \item In what ways do art historians conceptualize critical reflection in the context of human-AI collaboration?
    \item How do art historians realize critical reflection when interacting with a CV-based image retrieval tool?
\end{enumerate}  

We conducted semi-structured expert interviews and a software exploration with 12 art historians. Our work does not aim at creating representative or generalizable assumptions on human-AI collaboration. This is why we deem this sample size well-suited for our research question. By assuring relative heterogeneity within our sample (see~\autoref{sec:participants}) while at the same time limiting the sample to a moderate size, we set the foundation for us to capture the complexity and nuance contained within the data~\cite{BraunClarkeSampleSize}. 

In the following, we describe the use case of CV-based image retrieval. Following this, we detail how we recruited our participants, conducted the interviews and software exploration, and performed a reflexive thematic analysis (TA) of the data collected.

\subsection{Use Case: Computer-Vision-Based Image Retrieval}
\label{sec:UseCaseComputerVision}
Our work rests on a clear distinction between ready-to-use tools that support the \emph{retrieval of objects} and \emph{computational analysis} that promises to yield actual results that are a relevant contribution to a field of study. In short, the tools that we foreground are intended to be used by art historians to find objects of study - not to inquire their objects of study.
Our decision to focus on image retrieval is informed by the observation that, in the humanities, digital tools are often used during the exploration phase of a research process, for example, while searching in digitized collections~\cite{koolen_toolcriticism_2018}. This decision is also informed by work on human-AI collaboration in the context of qualitative research (see \autoref{sec:humanAICollab}), which has pointed out researchers' skepticism toward the applicability of computational approaches during analysis and interpretation. 
These findings are congruent with the skepticism regarding computational analysis in the humanities (see, e.g.,~\cite{da2019computational, dobson2019critical, Klinke_2016}). The actual art-historical analysis and interpretation of the resulting corpus that includes the computationally retrieved objects would still be up to the scholars themselves~(see, e.g., \cite{Klinke_2016}).

CV is already broadly available in web-based, commercial applications like Google image search or as a feature in the iOS Photos app. However, such commercial applications do not serve the needs of art historians in the sense of a ``research tool.'' The lack of context-specific tools is why current research in DAH explores CV as a means for image retrieval that also gives room to more complex understandings of visual similarity. For example, images can be considered in their entirety or by indicating a more specific search interest such as a specific motif or pose. This can be done, for example, by drawing bounding boxes around image elements (see, e.g.,~\cite{seguin2018replica, Ufer_image_retrieval, impett2020analyzing})\footnote{As pointed out above, our use case is focused on image \emph{retrieval}, not on image \emph{analysis and interpretation} as a means of scholarly knowledge production. Nonetheless, we also reference research that goes beyond this scope or is framed as enabling art historical \emph{analysis}. We do so in order to give a more comprehensive overview of research at the intersection of computer vision and art history.}. Such applications of CV help researchers to find artworks that - in their entirety - do not look similar to the input image but that do include similar motifs~\cite{Ufer_image_retrieval}. Despite the breadth of research on CV for art historical study purposes, the solutions are, as of yet, rarely made available to nontechnical users in ready-to-use tools~\cite{klic2022linked}. 

\begin{figure}[h]
\caption{Screenshot of the \texttt{imgs.ai} interface after performing a search with three ``positive'' and three ``negative'' images as input using the embedding ``poses''.}
\includegraphics[width=\textwidth]{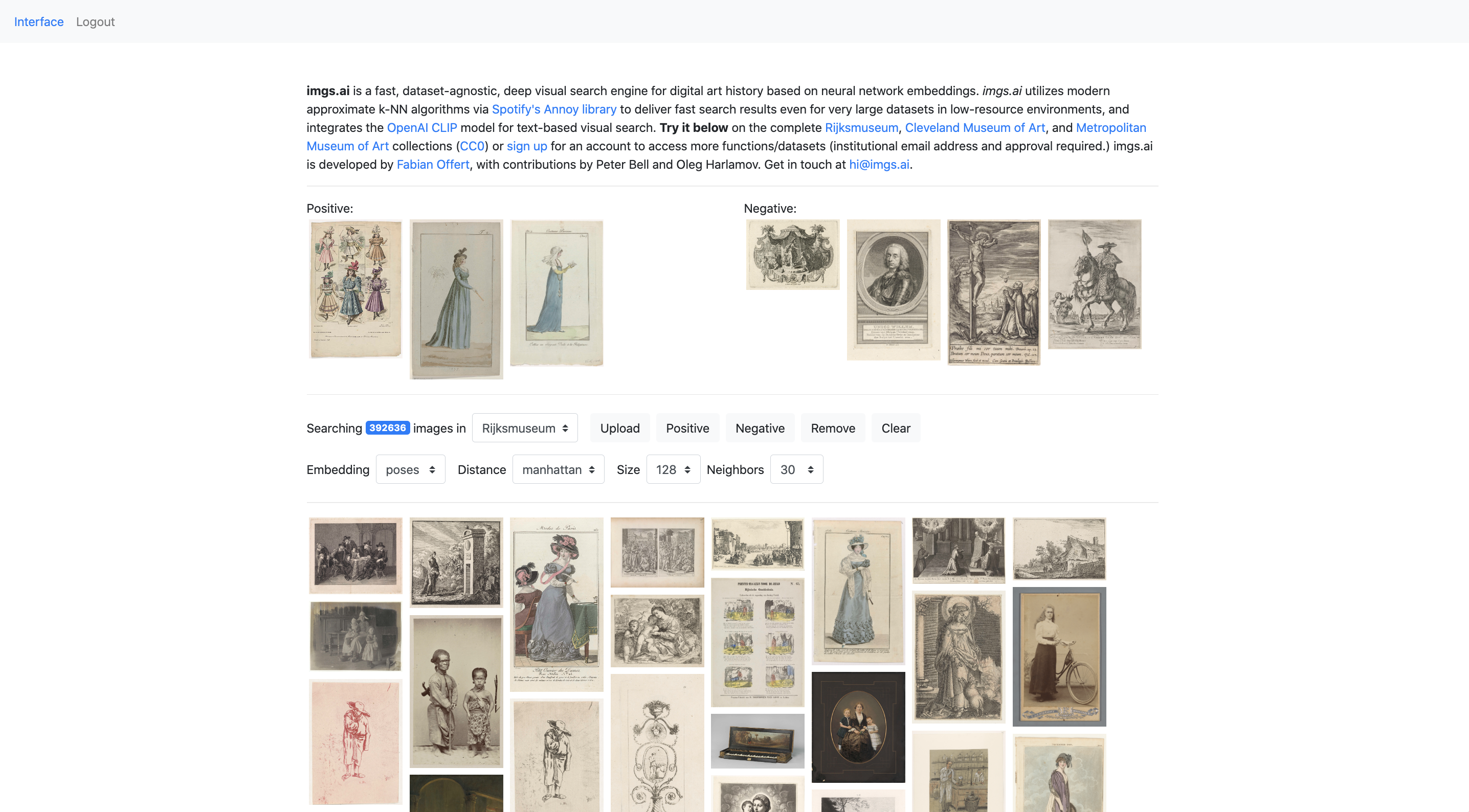}
    \Description[Screenshot of the \texttt{imgs.ai} interface after performing a search with three ``positive'' and three ``negative'' images as input using the embedding ``poses''.]{long description for visually impaired readers using a screen-reader}
    \label{fig:imgsai}
\centering
\end{figure}

One of the few examples of a ready-to-use tool that integrates CV for visual search on art historical image datasets is \texttt{imgs.ai}~\cite{imgsai2023}\footnote{Further information is provided on \url{https://imgs.ai/}.}. 
The web-based tool allows users to upload one or more images and perform a visual search on a selection of museum collection datasets. During their interaction with the tool, users can continuously refine their explorations or searches within a collection by using the results of one search as visual input for the next search (``re-search''). This is done primarily by selecting images from the tool's output of `similar' images and marking them as either ``positive,'' i.e., matching the user's search interest, or ``negative,'' which prompts another search. The tool also enables cross-collection searches. The tool's functionality is based on feature extraction. It enables users to interactively select different embeddings, i.e., compressed semantic descriptors for images, and a distance metric as a similarity criterion when performing their search ~\cite{imgsai2023}. The operation of the ready-to-use web-based search tool does not require advanced technical knowledge. 

We are not involved in the development of this tool but see \texttt{imgs.ai} as a typical example of a computational tool for humanities research. Therefore, we decided to use \texttt{imgs.ai} as part of our study during the software exploration (see~\autoref{sec:interviews}).  

\subsection{Participants}
\label{sec:participants}

Our research follows a qualitative approach and builds on expert interviews and a software exploration with art historians. We compiled a list of potential participants in preparation for the study. The prerequisites for inclusion on this list were that the candidates have at least one academic degree in art history and work in an art history-related profession. The initial selection was based on our professional networks. It was extended by researching art historians listed on institutional websites of universities, research institutes, museums, or other collecting institutions. We snowballed additional candidates during the first round of contacting participants. The final list included 35 potential study participants. All candidates were assigned a \textit{level of familiarity with computational approaches} ranging from `A - high' to `C - low'\footnote{In order to assign this level, we gathered information from their profiles on institutional or personal websites and from their publication history. Participants who have contributed significantly to computational tool development or list other software development experiences in their profile were grouped as `A - high.' Participants who work in the context of digitization efforts, for example, in museums or institutional collections, were grouped as `B - medium,' as were participants who have published in a journal or at a conference that can be counted as belonging to digital subfields of art history or the digital humanities. Participants who have no known record of either using computational approaches for their research or engaging theoretically with computational approaches were grouped as 'C - low.' We adjusted the assigned levels of familiarity in two instances based on the participants' self-reported experiences with computational approaches and tools during the interviews.}.
We contacted 17 potential participants from the initial list, of whom 12 accepted our invitation. We ensured that our final sample included all levels of familiarity with computational approaches. Thus, we contacted candidates successively, i.e., contacting `replacement' candidates depending on their level of familiarity. This measure was taken to assure relative heterogeneity in our sample and to reduce probable `skepticism' or `enthusiasm' bias towards computational approaches that might correlate with the level of familiarity. Three of the 12 participants were qualitatively grouped as `A,' i.e., high familiarity with computational approaches, five as `B,' i.e., medium familiarity with computational approaches, and four as `C,' i.e., low familiarity with computational approaches. 

\begin{table}[t]
    \caption{Study participants.}
    \centering
    \begin{tabularx}{\textwidth}{l>{\hsize=.6\hsize}L>{\hsize=1.1\hsize}LL}
        \toprule
        Pseudonym&Level of familiarity with computational approaches&Function&Professional context\\
        \midrule
        P1&B (medium)&professor&university / research institute\\
        P2&B (medium)&researcher / curator&museum~/~institutional~collection\\
        P3&A (high)&post-doctoral~researcher&university  / research institute\\
        P4&C (low)&post-doctoral~researcher&university / research institute\\
        P5&C (low)&professor&university / research institute\\
        P6&A (high)&researcher / curator&museum~/~institutional~collection\\
        P7&A (high)&doctoral~researcher&university / research institute\\
        P8&B (medium)&post-doctoral~researcher&university / research institute\\
        P9&C (low)&doctoral researcher&university / research institute\\
        P10&B (medium)&post-doctoral~researcher&university / research institute\\
        P11&C (low)&researcher / curator&museum~/~institutional~collection\\
        P12&B (medium)&researcher / curator&museum~/~institutional~collection\\
        \bottomrule
    \end{tabularx}
    \Description[Table showing the level of familiarity and the role of each subject]{Table showing the level of familiarity and the role of each subject}
    \label{tab:participants}
\end{table}

Seven of the 12 participants identified as women and 5 as men. All participants have at least a university degree (M.A.) in art history, often combined with another humanities degree. All participants work in an art history-related profession. Four participants work as researchers or curators at a museum or institutional collection, and eight participants work at a university or research institute, ranging from doctoral researchers to professors~(see~\autoref{tab:participants}). Eight of the participants were from Germany, the others from Switzerland. 

Our university does not have an institutional or ethical review board. To compensate for this, we have developed internal guidelines that set ethical and legal standards for our research group. These guidelines include sending all potential participants a thorough information sheet upon first contacting them, which allows them to determine whether or not they are willing to participate in the study. They were also informed that they would not be compensated for participating in our study. Upon acceptance of participation and prior to the interviews, participants received another detailed information sheet that informed them about the background of the study, detailed the process and how the interview will be conducted, and how the data collected will be used and stored.
The document, which is approved by our university's Chief Data Protection Officer, also informs them of their rights under the General Data Protection Regulation. 
The interviews were conducted remotely through the video conferencing software `Cisco Webex,' which was also used to record the interviews' audio and video. Participants were informed of the circumstances of the recording and consented to it. Cisco Webex is approved by our university's Chief Data Protection Officer. 
The audio and video recordings and the transcriptions are stored on an internal password-protected server and will be deleted after the completion of the project. 

\subsection{Expert Interviews and Software Exploration}
\label{sec:interviews}

The semi-structured expert interviews and software exploration were conducted in March and April 2022. The interviews were each planned to last 35 minutes, and the software exploration another 25 minutes. The average duration of both parts was about 70 minutes. All interviews were conducted in German. 

The first part, the semi-structured interviews, focuses on understanding our participants' research processes and work contexts. We were similarly interested in learning more about what roles the use of computational tools and collaboration play in their research. The interview script also included questions that encouraged the participants to share what potentials and affordances they associate with non-specified AI systems, especially systems that integrate CV. As laid out in \autoref{sec:introduction}, we propose critical reflection as a `meaningful' realization of human-AI collaboration and hypothesize that tools need to be intentionally designed in such a way that they support critical reflection. In order to be able to test our proposal and hypothesis, we considered it essential not to prime our participants by directly addressing critical reflection. In an iterative process, we carefully crafted the interview script accordingly. We only encouraged participants to expand on topics relating to criticality and reflection when they brought them up themselves. The interview questions are provided in~\autoref{InterviewProtocol}. 

When we contacted the participants, we informed them of the study procedure and that they are invited to use the freely available CV-based search interface \texttt{imgs.ai}\footnote{For more information on this tool, please refer to \url{https://imgs.ai/}.} during the software exploration. For this purpose, we asked them to bring a digital image with which they could perform an image retrieval task that is relevant to their research or current work. We asked the participants to share their screen, which allowed us to record and follow their interaction with the tool. It also enabled us to insert questions or provide help when needed. 

At the beginning of the software exploration, we only briefly introduced the tool and that the participants could perform image-based searches with it. The decision not to explain the tool's functionalities in greater detail before having the participants use it mimics how they would use such a tool as part of their everyday work. 
The use of search interfaces to web-based repositories, image catalogs, collection databases, or Google image search is not ``introduced'' and explained either. At least not beyond what the interface offers as guidance or through links to additional external information or documentation. Nonetheless, we ensured that all participants reached a sufficient understanding of the tool's functionalities and supported them when we noticed fundamental misconceptions about how the tool works and how they can operate it. Such additional support was only needed in two instances. Following the familiarization, we asked the participants to use the image they had brought to perform an image retrieval task relevant to their research or current work practice. We instructed the participants to verbally share their thoughts, observations, and questions that arose during the interaction, i.e., to think-aloud~\cite{Lewis_1982} while using the tool. We also encouraged them to verbalize their search interest and their interpretation of the tool's output. 

\subsection{Thematic Analysis}
\label{sec:thematic_analysis}

We analyzed more than thirteen hours of audiovisual footage. Our analysis was guided by Braun and Clarke's `reflexive thematic analysis' (TA) approach, which conceives analysis as an active creation of themes by the researchers at the intersection of data, analytic process, and subjectivity~\cite{Braun_Clarke_2006, BraunClarkeRTA}. Our choice of reflexive TA was motivated by the acknowledgement of subjectivity as a central feature and requirement within this analytic approach. Hence, it allows us to embrace our interpretative lenses that are informed by our disciplinary backgrounds\footnote{We both work in a research group in the field of HCI, with a particular focus on machine learning (ML) technologies. We understand and acknowledge that our academic preoccupation with ML in terms of reflection and interpretability, as explored in CSCW and HCI research, shapes the study presented in this paper. Our work is informed by our complementary academic backgrounds; the first author has a background in cultural theory, art history and (digital) media studies, and the second author in collaborative computing and HCI.}. 
Additionally, we deem the reflexive TA approach flexible enough to be applicable not only to our interview data but also to the data gathered from the software exploration. The emphasis that is put on the researchers' interpretation of the data when identifying patterns~\cite{Braun_Clarke_2006, BraunClarkeRTA} encouraged us to consider the participants' interaction and behavior during the software exploration as another layer of information. 

We transcribed all interviews, including the recordings of the software exploration. In addition to transcribing what the participants said during the `thinking-aloud' software exploration, we annotated the transcriptions with what the participants were doing, i.e., how they interacted with the tool, to contextualize the spoken word. After familiarizing themselves with the data, the first author initially coded all 12 transcripts, which served as a basis for several rounds of iterations on the coding. This phase included exchanges between both authors, during which we revised and reassigned our codes and shared our observations. Our inductive coding was grounded in the data, but we acknowledge that our theoretical assumptions and knowledge shape the analysis. Instead of seeking consensus on meaning, our exchanges and discussions gave room to subjective readings and interpretation of the data, as outlined by Braun and Clarke as one of the characteristics of a coding approach that is in line with reflexive TA~\cite{BraunClarkeRTA}. 

We coded and analyzed the transcripts in two batches. We analyzed the interview data (first part) separately from the data gathered during the software exploration (second part). This allowed us to compare the two parts and relate them to the dimensions of our research question, i.e., our four sub-questions (see~\autoref{sec:methods}).
Following the coding phase, we generated a set of candidate themes that we organized into a `thematic map.' We used this map in the subsequent steps of our analysis and discussions to organize and restructure the themes we had developed based on the data. From our candidate themes, we selected those that we considered being the most significant for our research question. We carefully translated selected quotes from German to English for integration them in this paper and intensively discussed gradual shifts in meaning that emerge from such translations. 

\section{Findings}
\label{sec:findings}

Through our analysis, we developed eight themes. In the following, we present the themes structured into four main sections corresponding to our sub-questions in~\autoref{sec:methods} (see~\autoref{tab:findings}). The first section contains themes that relate to the potentials and affordances participants ascribe to human-AI collaboration. The second section lays out themes that relate to how these potentials and affordances surfaced during the participants' interaction with the CV-based image search tool \texttt{imgs.ai}. The themes in the third section summarize how participants conceptualize critical reflection in the context of human-AI collaboration. Lastly, the themes presented in the fourth section illustrate how critical reflection is realized by the participants during their interaction with a CV-based image retrieval tool. Due to the complexity and richness of our data, our analysis led to the creation of a larger number of themes. However, we only foreground the themes that directly relate to our sub-questions. 

\begin{table}[thb]
    \centering
    \caption{Overview of themes organized in four sections.}
    \begin{tabularx}{\textwidth}{LL}
        \textbf{\autoref{potentialsaffordances}: Potentials and Affordances Ascribed to Human-AI Collaboration}&\textbf{\autoref{SurfacingPotentials}: Surfacing of Ascribed Potentials and Affordances during Interaction with CV Tool}\\
        \toprule
        Computational tools are not able to produce reliable (research) results automatically and should not be considered for this purpose&CV tool is expected to produce reliable `results' without further interaction
        \\\midrule
        Computational tools and approaches have the potential to influence epistemology&Tool output is not considered purposeful when it does not align with established epistemologies\nextrow
\nextrow
        \textbf{\autoref{Conceptualization}: Conceptualizations of Critical Reflection in the Context of Human-AI Collaboration}&\textbf{\autoref{CRinteraction}: Realization of Critical Reflection during Interaction with CV Tool}\\
        \toprule
        Critical reflection is considered a foundation of research and is expected to be applied in human-AI collaboration&Source criticism and reflection on methods were not fully realized during the interaction
        \\\midrule
        Participants' expectation to apply critical reflection is directed at themselves&Participants expect tool to enable critical reflection during interaction
    \end{tabularx}
    \Description[Table showing the themes as laid out in Section 4.1 to 4.4]{Table showing an overview of the themes as laid out in Section 4.1 to 4.4}
    \label{tab:findings}
\end{table}

\subsection{Potentials and Affordances Ascribed to Human-AI Collaboration}
\label{potentialsaffordances}

The participants were made aware of our research into human-AI collaboration through the information sheet we sent them in preparation of our study (see~\autoref{sec:methods}). We repeated this information at the beginning of the interview (see~\autoref{InterviewProtocol}). However, in line with our understanding of human-AI collaboration as an \emph{analytical lens}~(see~\autoref{sec:introduction}), we did not prompt the participants to detail their understanding of what, in their opinion, constitutes `human-AI collaboration.' Instead, we asked participants about their experiences with computational tools and approaches and what role 
such tools and approaches play in their research and art historical work. We also encouraged the participants to share their ideas and assumptions regarding non-specified AI, particularly CV, even when they did not report any experience in using such technology. In this regard, participants also speculated about how AI might affect art-historical research and work practice. In the following, we highlight two themes that relate to the potentials and affordances that the participants ascribed to human-AI collaboration.

\subsubsection{Computational tools are not able to produce reliable (research) results automatically and should not be considered for this purpose.} 
\label{tool_limitations}

Participants were generally open to using computational tools and exploring computational research approaches. In the context of art history, they saw a particular potential in CV and were interested to see how this technology might influence their research or work with collections. By and large, they conceived of computational tools as something that should only \emph{support} their epistemic process but not replace it. However, several participants feared that computational tools could be misinterpreted as being able to produce research results by themselves.
\begin{displayquote}
\textit{``It is important to realize that a tool cannot replace the research process, that it is just a tool - like a kind of notebook or database - that is used and cannot deliver the results. I don't think that's entirely clear, as trivial as it sounds, because there is a great hope that computational tools will not only support a research process but will themselves produce a research result.''}
(P7)
\end{displayquote}
Many participants saw potential in delegating certain clearly defined tasks to a computational tool. However, they considered it a requirement that humans are included in the process. They mentioned, for example, that it is important that humans ``check'' the output of a CV tool because they do not expect the tool to perform reliably enough or be accurate enough in terms of what concepts and levels of `similarity' are needed in art history and when working with original objects.
\begin{displayquote}
\textit{``I think you still have to do a cross-check since computer vision can't solve everything 100 \%. With prints, there are sometimes minimal differences in the editions, so two images might only look like two identical prints at first glance. Then you have to take a closer look [for example, at the material].''}
(P11)
\end{displayquote}
Participants mostly conceived computational tools as an addition to their research and work practice. They considered it essential to combine computational approaches with ``conventional,'' i.e., non-computational methods and research practices.
\begin{displayquote}
\textit{``But maybe it would be enough to say, okay, you get to a certain level and then continue conventionally. For me, it is not about the question that we are replaced and fully automatically have all the answers. I don't want that at all. But maybe I just want to be stimulated to think further in another direction and conduct a preliminary investigation [using a computational tool] for that - or I am at a certain point and then want to use it. This combination is actually rather what I would associate with it.''}
(P1)
\end{displayquote}

\subsubsection{Computational tools and approaches have the potential to influence epistemology}
\label{sec:influence-epistemology}

Participants frequently pointed out that computational research approaches and tools, especially those that integrate `AI,' have an influence on epistemology, i.e., they influence what questions can be asked, what methodology and knowledge are considered, and how objects of studies are approached. This influence on epistemology was generally considered productive and welcomed. However, for it to be seen as productive and welcomed, it had to remain in line with the previous theme: computational research approaches and tools should not be considered as being able to produce research \textit{results} by themselves. A welcomed influence of computational approaches and tools is, for example, that they enable a researcher to come up with new research questions. This, in turn, might lead to a reconsideration or change in the epistemic process. 
\begin{displayquote}
\textit{``When we work with computational tools, I at least try to make sure that I'm aware that it brings up different questions. And that's also okay; that's also what makes it interesting.''} 
(P12)
\end{displayquote}
\begin{displayquote}
\textit{``And I believe that this is exactly the potential - whether it is computer vision or other AI-supported research: that one comes up with questions one had not thought of before.''} (P1)
\end{displayquote}
%
%
When talking specifically about CV, some participants saw the potential that the technology could contribute to counteracting canonical structures\footnote{The canon can be understood as `biases' that are ingrained in collections, for example, in regards to what objects are considered artworks and are included in museums of art versus what objects are considered as `artifacts' and included in ethnological museums.}. Canonical structures, for example, influence what material will be considered `relevant' in certain research areas.  
\begin{displayquote}
\textit{``From my point of view, [CV tools] could help in particular in terms of a similarity search. For example, they can provide access to and insights into other genres of images that one would not find in existing canonical catalogs - so, in a sense, they could support a `visual science' [Bildwissenschaften] approach. In this way, they can also certainly bring forward new comparabilities and new research questions.''}
(P6)
\end{displayquote}
\begin{displayquote}
\textit{``When I'm trying to question my own view - so, when I, for example, search for representations of factories and immediately think of Menzel, Meyerheim, Karl Blechen, and so forth; the canon - then I just try a search with some CV tool, even if I don't really know how it's trained, and then I just find something else.''}
(P8)
\end{displayquote}
%
%
%
Participants generally valued the potential of CV to extend established text-based approaches to image retrieval. Text-based access to objects was, for one, problematized in terms of reproducing and manifesting established and canonical interpretations (see \autoref{sec:CV_ArtHistory}). For the other, participants valued the potential of CV to enable retrieval that is in line with thinking visually.
\begin{displayquote}
\textit{``[Computer vision holds the potential] that one develops a desire to discover art-historical data and images differently and perhaps not always through text. We are so used to always entering search terms, or we have these thesauri and that helps us to generate structured queries. But people also think in images.''}
(P4)
\end{displayquote}
Correspondingly, some participants conceived CV as potentially influencing their own evaluation of artworks. This was mentioned, for example, in terms of what objects they consider as being `similar' (and, thus, as possibly relevant for later analysis). Text-based access to art-historical objects is confined to established orders of knowledge and canonical interpretations of artworks (see~\autoref{sec:CV_ArtHistory}), which do not always correspond with an apparent visual similarity. Shifting to an image-based retrieval mode was valued as enabling searches based on the object's visual similarities. Participants appreciated that this enables them to include objects in their research that they would not have considered `similar' based on a text-based search. One participant mentioned that CV, in this sense, would present them with ``visual evidence'' that they had not considered themselves:
\begin{displayquote}
\textit{``You suddenly get visual evidence somehow. That's really nice sometimes, that you see at first glance: I'm only now becoming aware of this similarity. And that also includes surprising results! So that's where I see the great advantage of an image-based search, that you actually gain this surprising visual evidence.''} (P3)
\end{displayquote} 
\subsection{Surfacing of Ascribed Potentials and Affordances during Interaction with CV Tool}
\label{SurfacingPotentials}

We seek to understand how the ascribed potentials and affordances that we detailed in \autoref{potentialsaffordances} surface during the interaction with a CV-based tool. For this purpose, we analyzed the data gathered from the software exploration and related it to our analysis of the interview data. 
In the following, we detail two themes that indicate misalignments between the potentials and affordances that participants ascribed to CV during the interviews and how they interacted with or verbally reacted to the tool. This includes, for example, instances of misalignments in expectations. We interpret those misalignments as indicators that the provided information in the tool was insufficient or unsuited to manage expectations regarding the tool's potentials and affordances. Hence, we address divergences as indicators for needed improvements in future tool design (see~\autoref{sec:designimplications}).

\subsubsection{CV tool is expected to produce reliable `results' without further interaction.}

During the interviews, participants generally conceived computational tools as unable to produce reliable (research) results by themselves. This conception coincided with the implication or explicit acknowledgment of the need to check, interpret, and revise tool output, i.e., to iteratively and critically interact with a tool and its output. Participants also were aware of the limitations of CV in regards to being able to capture their domain-specific concepts of similarity and levels of detail~(see Section~\ref{tool_limitations}). However, during the software exploration, we observed that some participants expected the tool to present them with reliable `results' for complex\footnote{In terms of `complexity,' we refer to searches that are contingent on high levels of detail in object similarity that would require extensive contextual knowledge, for example, expert knowledge of non-visual material properties, or a high degree of visual literacy, for example, in terms of identifying differences in print techniques.} search interests without further interaction. In such cases, participants showed some reluctance to interact with the tool, for example, in regards to refining the parameters or input, and to interpreting how the tool output related to their input image. 

We interpret such instances as overconfidence in the tool's abilities and expected level of precision. We acknowledge that these effects might have been exacerbated by the framing of the tool, which has to be factored in as a layer of information contained in the interface (see~\cite{drucker2014graphesis})\footnote{The information provided in the interface presents the tool as ``a fast, dataset-agnostic, deep visual search engine for digital art history based on neural network embeddings.'' We believe that this framing as a tool ``for digital art history'' has to be factored in as having an effect on the participants' expectations and as presumably contributing to such overconfidence, i.e., increasing their expectations regarding the 'precision' of the tool compared to the level of precision that they are familiar with in `general purpose' image search such as Google image search.}. One participant, for example, had selected a 17th-century engraving of Hercules (a depiction of a very muscular man) and wanted to retrieve other depictions of muscular men, i.e., other depictions of Hercules or related mythical figures. The tool output, however, also included engravings depicting muscular women and mythical figures with androgynous bodies. To an untrained human eye, those images of muscular bodies could be considered `similar' to the input image. In this case, the level of similarity in the tools' output was not aligned with the expected precision and specificity of the search interest.

\begin{displayquote}
\textit{``[$\ldots$] Although I said that women are not my search interest, quite a few women appear here. That would annoy me now, of course.''}
(P11)
\end{displayquote}

Another participant had uploaded a 16th-century print depicting cherubs (putti) drinking together. When the system returned images of drinking adults, they felt that the tool had misunderstood their input. 
\begin{displayquote}
\textit{``Well, that means that it has not grasped that I'm actually looking for five small children dousing their noses. Instead, I am now presented with many gentlemen drinking alone and an elderly lady. So this is not very helpful for my search.''}
(P5)
\end{displayquote}
%
One feature of \texttt{imgs.ai} is that it allows users to select images from the tool's output as ``positive,'' i.e., in line with their search interest, or as ``negative,'' which triggers a new search. We ensured that all participants had familiarized themselves with this basic functionality during the initial exploration phase (see~\autoref{sec:interviews}). Some participants valued the back-and-forth interaction with the tool's output, i.e., that they could refine the results based on what they considered `similar' to their input image and also adapt parameters such as the embedding selected. However, in some instances, interacting with the tool in a `collaborative' sense was perceived as a burden. 
\begin{displayquote}
\textit{``There is a lot of noise among these results. [Interviewer suggests that P7 refines their search by entering those images as `negative.'] Okay, well, then let's do that. Well, one is more or less spoiled by Google image search, because there, one doesn't need to tinker with the search query at all [$\ldots$].''}
(P7)
\end{displayquote}
%
%
%

%
\subsubsection{Tool output is not considered purposeful when it does not align with established epistemologies} 

As summarized in Section~\ref{sec:influence-epistemology}, participants ascribed and valued the potential of computational tools and approaches to bring forth new questions and epistemic perspectives. However, we often observed divergences from this conception during the software exploration. Most participants approached the tool with a `conventional' search interest that aligned with established epistemologies. 
%
Commonly, search interests corresponded to a clearly defined era and culture of origin that the participants were interested in. These served as clear-cut exclusion criteria when participants judged the tool's output. When the tool presented them with suggestions of visually similar images, many participants evaluated the suggestions solely based on whether or not they were in line with their exclusion criteria. 
\begin{displayquote}
\textit{``Art-historical research is very specific [$\ldots$]. There is either the cultural context or the historical context, or other relations. So a search has to relate to that in some way. The idea that an image is just somehow visually similar to another one that comes from a completely different context, that doesn't fit well into art-historical research''.}
(P7)
\end{displayquote}
However, we also observed instances where participants started with a `conventional' search interest that they then developed further during their interaction with the tool. 
Participants who had emphasized during their interviews that they valued the potential influence that computational approaches might have on epistemology tended to be more open toward this aspect during the software exploration. This was, for example, the case with P1. They had initially dismissed the tool's output as not valuable to their search interest, which, in this case, was to find images that depict the birth of the Virgin Mary. 
While interacting with the tool, however, their openness toward epistemological changes that they had expressed during the interview eventually surfaced during their interaction with the tool.  
\begin{displayquote}
\textit{``So, the most interesting to me is the last embedding, because it brings me to new questions. The others are all conventional - iconographic or generic - results. [...] But here [after selecting the embedding `raw'], I would immediately come up with new questions that I would pursue.''}
(P1)
\end{displayquote}

\subsection{Conceptualizations of Critical Reflection in the Context of Human-AI Collaboration}
\label{Conceptualization}

As laid out in \autoref{sec:methods}, we carefully crafted the interview script in such a way that it would not prime the participants on the aspect of critical reflection (see \autoref{InterviewProtocol}). Nonetheless, in line with our proposal for `meaningful' human-AI collaboration, participants regularly emphasized the importance of critical reflection. In the following, we highlight two themes relating to participants' conceptualizations of critical reflection. 

\subsubsection{Critical reflection is considered a foundation of research and is expected to be applied in human-AI collaboration}

Participants generally considered critical reflection a foundation of research and related it to methodological scrutiny. 
In some instances, the importance of critical reflection in art-historical research contexts was motivated with reference to the ``methodological problem'' of art history and other humanities disciplines, i.e., that they are not considered ``scientific,'' as P9 argued:
\begin{displayquote}
\textit{``Art history has always had - since the beginning of the discipline - the methodological problem that it is not a natural science, and the methods... well, they fray at the edges and are sometimes quite wishy-washy, which is also something that art history is always accused of. [...] But in this respect, my personal view of scholarly work in art history is that one should always reflect on oneself and one's theses.''}
(P9)
\end{displayquote}
%
%
Participants often mentioned critical reflection when we asked them which competencies they considered most important in the context of using computational tools and approaches in art-historical research. Participants commonly conceptualized critical reflection in relation to two central concepts in humanities research: \emph{source criticism} and \emph{reflection on methods}.
Source criticism was, for one, implicitly and explicitly referred to as the ability to judge the reliability of a source and the provider of information or data. This ability included, for example, knowing which institution provides ``rich'' metadata with their digitally published collection. 
\begin{displayquote}
\textit{``A trustworthy data source is, for example, the British Museum's database because I just know that their collection is very well indexed and that they already started to fill their database systems 20 years ago.''}
(P10)
\end{displayquote}
Participants also emphasized that it is important to be aware of the historicity of collection data. In this regard, they consider it necessary to critically reflect on the madeness, situatedness, and incompleteness of data and to understand selection criteria and biases in collections. 
\begin{displayquote}
\textit{``Who wrote the data? Are these perhaps simply not `data' at all, but only what was already written 100 years ago in catalogs? There's this aura of newness that surrounds all this technical stuff [...] - and I try to teach my students how old approaches are actually just translated. For me, it's important that they critically reflect and understand what kind of information they're working with.''}
(P8)
\end{displayquote}
%
\begin{displayquote}
\textit{``If we now integrate large masses of images with old descriptions into databases without thinking about what kind of descriptive criteria we have, we will only reproduce those old descriptions. I do wonder whether we are simply perpetuating racism or hidden narratives, which then simply live on in the data.''}
(P5)
\end{displayquote}
In this context, participants were concerned that computational approaches might be, or are, perceived as producing reliable, `objective' and quantifiable truths, which makes it harder to question, scrutinize or critically reflect on them. Participants elaborated on this concern, for example, in terms of reproducing canonical structures and biases in collection databases. 
\begin{displayquote}
\textit{``If I have a huge database that promises to serve me with the metadata of 10,000 images and answer all my questions - then I wonder who still questions this [...]. I sometimes have the feeling that if it's in a digital format, you can see even less behind it, and maybe you can't even say: `Funny how much data there is, but why is it mainly white European men who are represented in the data?'''}
(P5)
\end{displayquote}
%
%
In addition to such instances of ``source criticism'' and reflection on data, conceptions of reflection on methods were particularly present when participants considered the effects of computational approaches on the epistemic process. They emphasized the need to understand how the results produced with the help of a computational tool came about. Participants considered this a prerequisite for being able to reason with and about the tool's output in a critical-reflective manner. 
%
%
This aspect was often brought up when participants talked about non-specified AI. 
%
\begin{displayquote}
\textit{``Well, of course, you need as much transparency as possible regarding accounting for the process. And, I mean, as an art historian, you write essays when you argue. The machine, if I see it correctly, eventually spits out a result. You can then write about the programming of the whole thing or how this `ingenious brain' was conceived, but understanding how it actually reaches a result would of course be the interesting thing about it.''}
(P9)
\end{displayquote}

\subsubsection{Participants' expectation to apply critical reflection is directed at themselves}

As summarized in the previous theme, participants emphasized the importance of critical reflection and expected it also to be realized in human-AI collaboration. This expectation is directed at themselves or fellow art historians or humanities scholars. They consider it their own responsibility to acquire the skills needed to realize critically-reflective usage of computational tools or data. 
\begin{displayquote}
\textit{``We always need a critical mind, not only as scholars. [...] Just because a tool spits out a result, you still have to think along and ask yourself: does this make sense? I need to have a critical awareness - despite the very high accuracy that these tools offer - I have to remain vigilant. I would consider that to be an important competency. [...] Of course, you must first be trained a bit in order to master this.''}
(P11)
\end{displayquote}
Participants consider it important to teach students how to work with data critically, which could build on existing competencies in the humanities regarding the handling of sources and information. They emphasize that it should be part of art-historical curricula to teach students how to handle data, understand their origin, and what research questions could (and could not) be answered based on them. However, in regards to reflecting on methods, i.e., reflecting on how a CV tool produces a result, P7 states that this is not as easy to achieve:
\begin{displayquote}
\textit{``This is, of course, more difficult, because these computer vision systems are a black box, or they somehow present themselves as such.''}
(P7)
\end{displayquote}
Only one participant (P6) explicitly pointed out that the realization of critical reflection in human-AI collaboration should not be delegated to art historians without prerequisites: 
\begin{displayquote}
\textit{``On the one hand, this is a competency that we [art historians] have to acquire. On the other hand, I expect that all software manufacturers meet the requirements of transparency, so to speak. What are the data sources? How does it work?}''
(P6)
\end{displayquote}

\subsection{Realization of Critical Reflection during Interaction with CV Tool}
\label{CRinteraction}
During the interviews, participants consistently emphasized the importance of critical reflection when working with computational tools. This importance was also mirrored during the software exploration, where participants showed eagerness to realize such a critical-reflective approach. In many instances, participants were able to critically reflect on the mediality of the digital reproductions, for example, in regards to how the digital reproduction of objects influenced the visibility of material details. Participants also applied their competence regarding source criticism. This competence was realized by accessing detailed information about an image by clicking on the link in the tool's context menu, which would open the object's record entry in the institutional collection database in a new browser tab. 

However, as we detail in the following, we regularly observed divergences from the aspiration to interact with a tool in a critical-reflective manner. These divergences did not occur due to a lack of eagerness or capability on the side of our participants. Instead, we interpret this as support for our hypothesis that computational tools need to be intentionally designed in such a way that they actively scaffold and support critical reflection.

\subsubsection{Source criticism and reflection on methods were not fully realized during the interaction}
\label{reflection_notrealized}

Participants commonly applied their competencies in ``source criticism'' by scrutinizing the provided data and the source (see above, \autoref{CRinteraction}). Although some were able to extend their competencies to the intentional selection of a collection on which to perform a search, most participants started their search with the pre-selected collection\footnote{This observation is a testament to the influence of default settings, which has been extensively studied in a range of contexts and is known to have a considerable effect on selection~\cite{jachimowicz_duncan_weber_johnson_2019}.}. In such cases, the participants did not consider whether or not the default collection was likely to hold objects that would correspond to their specific search interest, which sometimes led to initial negative evaluations of the usefulness of the tool. 
The lack of consideration of the selected collection could be easily overcome when participants were made aware of their unintentional adherence to the pre-selection and encouraged to consider searching in another collection that better matched their search interest. 

Participants were likewise eager to reflect on the underlying methods of the CV tool. However, they frequently and consistently mentioned that the lack of transparency and explanations made it hard for them to scrutinize the tool's output, which also limited their ability to interact with the tool in a `meaningful' manner.
\begin{displayquote}
\textit{``I do not have the feeling that I [$\ldots$] understand the effects that the parameters have on the result. That is not intuitively recognizable to me, and I would like to know that before I operate the tool. So to me, this is just like operating a bread-cutting machine without knowing where I can adjust it to cut wider or narrower. This is not being explained and that's why I say, okay this is some kind of image recognition on some level. But [$\ldots$] this is not really understandable or verifiable and I would need an explanation of how [the embeddings] work.''
(P6)
}
\end{displayquote}
In light of the lack of explanations, some participants fell back on their competency as art historians and applied a hermeneutic approach to interpretation. That means that they approached the tool's output in an interpretative manner and tried to understand the underlying principles of similarity encoded in the different embeddings by way of visual analysis and comparison of the outputs.  
%
%
\begin{displayquote}
\textit{``[After changing the embedding] it shows me less drawings. The shape is again more strongly represented, the round shape. [$\ldots$] There are many prints among the results. Images that I would spontaneously classify in the direction of woodcut, copperplate engraving, which are generally techniques that require a very controlled way of working, which of course influences the character of a work.''}
(P12)
\end{displayquote}
However, in several cases, the tool's lack of transparency negatively influenced the participants' willingness to engage with the output. In such cases, participants dismissed the tool's output as illogical or as `random' without trying to interpret what kind of `similarity' the images in the search result had in common. 
This was the case with P2: they used a painting depicting fish as input. The tool's output also included images of objects relating to the ocean or other marine species. P2 was not able to make sense of why the search results were not limited to images of fish: 
\begin{displayquote}
\textit{``So, I don't really understand why it displays these images as results, because they are very different. So we now have some images with fish or fish-like objects. But there is no logic behind it.''}
(P2)
\end{displayquote}
\subsubsection{Participants expect tool to enable critical reflection during interaction}
\label{crtoolsupport}
During the interviews, participants directed their expectation regarding the realization of critical reflection in the context of human-AI collaboration at themselves; they formulated it as an expectation that art historians and other scholars should strive after, for example, by acquiring computational knowledge or data literacy. 
By contrast, during the software exploration, they formulated this expectation predominantly as something that should be enabled by the tool \emph{during} interaction. 

The tool's interface includes a short descriptive text that links to GitHub repositories and documentation on some of the libraries or models implemented (see \autoref{fig:imgsai}). Three of the participants (P6, P8, P10)  clicked on these links and accessed the documentation and secondary sources of information. 
However, even those participants that accessed the documentation and secondary sources of information did so only briefly, i.e., they opened the links in new tabs and returned to interacting with the tool after a short glance at the documentation. Even though only three participants accessed the secondary sources of information, most participants were looking for more details on the technical parameters of the tool and assumed that a more refined version of the tool would include such information. 
\begin{displayquote}
\textit{``This `distance' is unclear to me. And what this embedding with the `vgg' is about is also unclear to me. But it will be explained in the final version, I suppose. [...] I imagine that there will be an `i' button for info and I will be able to access explanations about what the settings mean and what influence they have.''}
(P5)
\end{displayquote}
As pointed out in \autoref{sec:introduction}, we are skeptical about the effectiveness of linking to technical documentation and the appropriateness of such information for nontechnical users, which we see confirmed by our observations during the software exploration. In several instances, not being able to understand the tool's parameters resulted in the participants assuming that they were not supposed to alter them, let alone critically reflect on what effect they might have. This highlights the need to improve the design of tools so that they enable critical reflection \emph{during} usage and in a user-appropriate manner.

\section{Discussion}
\label{sec:discussion}


Our research is guided by the question of how we can enable `meaningful' human-AI collaboration. We explore this question by foregrounding a genuinely human capability -- critical reflection -- which we situate in the context of humanities research practice, specifically, art historical image retrieval. We conducted a qualitative interview study and software exploration with 12 art historians. Through our study, we developed a better understanding of the participants' ideas and assumptions regarding human-AI collaboration and associated concepts of critical reflection. By extending the expert interviews with a ``think-aloud'' software exploration, we were able to relate the findings from the interviews to how the participants interacted with a CV tool. Based on a reflexive TA on the data, we derived themes (see~\autoref{tab:findings}) that inform our proposal of design implications for critical-reflective human-AI collaboration. In the following, we highlight our key findings before we provide the design implications.

\subsection{Key Findings}

All interviewees conceived computational tools as a valuable addition to their research practice. Participants clearly displayed conceptions of computational tools that evoke instances of collaboration rather than automation, which echoes existing research into human-AI collaboration in qualitative research contexts (see~\cite{Jiang_serendipity, chen_ambiguity, Baumeretal:2020:Topicalizer}). Interviewees ascribed and valued the potential that AI might influence epistemology in art history. This included the premise that CV, in particular, might help to overcome canonical structures. Participants also indicated that CV could broaden scholarly access to objects by not solely relying on textual encodings of 'expert' knowledge during image retrieval. Such stances parallel critical reflective engagements with the hegemonic effects of text-based information systems (see~\cite{Rawson_AgainstCleaning, Waterton_STS_Archive, Feinberg_Database_CriticalDesign}). 

Participants brought up and elaborated on the importance of critical reflection throughout the interviews and considered it a foundation of (humanities) scholarship. Source criticism was, for example, referred to as a central research competency. Participants emphasized that this competency also needs to be applied when dealing with digitized sources, for example, when working with digital images and other art historical data. This aspect is already acknowledged within the CSCW and HCI community by calling for a more reflexive data practice (see~\cite{Scheuerman_etal_2021}) or, more generally, a reflexive documentation practice (see~\cite{Miceli_etal_2021}). 
Additionally, participants stressed that they consider it crucial that the output of computational tools, especially when integrating AI, can be scrutinized, which evokes instances of reflection on methods. Such scrutinization is especially acknowledged in research on human-centered explanations that help people to understand the results of AI methods (see, e.g.,~\cite{Chromik_Butz_2021, Fan_Yang_Yu_Liao_Zhao_2021}). In the context of working with data or using computational tools, participants generally expected that they themselves, and their discipline as a whole, should realize a critical-reflective practice. 
However, participants shared their concern that computational approaches and tools might be, or are, perceived as producing reliable, `objective' and quantifiable truths, which makes it harder to question, scrutinize or critically reflect on them. This concern is mirrored by research that shows that people often over-rely on results provided by AI (e.g.,~\cite{Buccinca_Trust_Think2021}). 

Through our reflexive thematic analysis, we identified divergences between the participants' aforementioned conceptions and how they interacted with and reacted to a ready-to-use CV tool during the thinking-aloud software exploration. 
Contrary to what they posited during the interviews, participants often expected the tool to present reliable `results' for complex search interests more or less automatically. 
During the interviews, participants valued the potential of computational tools to allow them to explore new epistemologies. During the software exploration, we observed that participants approached the CV tool with search interests that were very much in line with established art-historical epistemologies. 
In some cases, participants were able to reflect upon this during their interaction with the tool and subsequently realigned their expectations and evaluation of the tool's output. 
These findings indicate that tools need to guide users in translating theoretically formed conceptions into practices of use.

Participants were generally eager to reflect critically on how the tool's output came about. However, the tool's lack of transparency and interpretability was frequently criticized as hindering them from realizing their own expectations in this regard. 
That this requirement was not met by the tool was interpreted as a sign that the tool is a `prototype.' 
At the same time, incomprehension sometimes prompted participants to engage in interpretation (see~\cite{baumer2015reflective}). In the absence of explanations regarding the underlying algorithms and the different concepts of `similarity' encoded in the embeddings, some participants fell back to their competency in hermeneutic interpretation when trying to make sense of or derive meaning from the tool's output. 
The option to test different embeddings and observe their effect on the tool's output encouraged some participants to reflect not only on the encoded concepts of similarity but also on the variety of notions of similarity ingrained in different art-historical approaches to analysis and interpretation. Such attempts at making sense of the algorithms and relating them to their own mental models and understandings of different concepts of similarity evoke the notion of the ``algorithmic imaginary'', as put forward by Bucher, i.e., ``the way in which people imagine, perceive and experience algorithms and what these imaginations make possible''~\cite{Bucher_algorithmic_imaginary2017}. 

Our findings confirm that art historians are eager to apply their competencies of criticality and reflection to the use of computational tools. However, we observed that they were not able to fully realize this eagerness and aspiration while interacting with a CV tool. We interpret such divergences as supporting our hypothesis that computational tools need to be intentionally designed in such a way that they actively enable and support critical reflection. Only then can humanities researchers such as art historians adhere to their own expectations of scrutiny and interact with computational tools in a `meaningful' manner. 

\subsection{Design Implications for Critical Reflective Human-AI Collaboration}
\label{sec:designimplications}

Our hypothesis resonates with Slovak et al.'s understanding of reflection as a process that needs to be scaffolded: we cannot assume that ``the ability to reflect is a trait that can be readily triggered by providing the relevant information''~\cite{Slovak_ReflectivePracticum}. Hence, Slovak et al. suggest ``scaffolding reflection within experience''~\cite{Slovak_ReflectivePracticum}. Our findings also echo this suggestion: Participants did not formulate their expectation that a tool should support critical reflection \textit{during} interaction with it until they engaged in a real-world image retrieval task with a ready-to-use CV tool. 
Taking this into account, we suggest four design implications for critical reflective human-AI collaboration.

\subsubsection{Supporting reflection on the basis of transparency}
\label{sec:DItransparency}

It is a recurring premise that art historians do not collaborate with colleagues and other experts~\cite{Brosens_SlowDigital2019} and consider art historical research a ``solitary endeavour''\cite{zorich2013digital}. Our findings could not confirm this assessment. Interviewees highly valued collaboration with colleagues from disciplines other than art history and the integration of multiple perspectives. 
Hence, the approach to take inspiration from human-human collaboration when designing human-AI collaboration (see~\cite{Wang_HHColltoHumanAIColl}) seems applicable in this case. 

Our interviews showed that art historians 
are able to reflect upon what expertise is needed for a given task at hand and initiate collaboration with colleagues and other disciplines accordingly. In cases where differing opinions stood in confrontation, they could factor in their collaborators' or colleagues' arguments and positions and either dismiss the suggestion or include it in their decision-making. The basis for such critical-reflective engagement with suggestions that ran contrary to their own assessments or findings was that they could understand the reasoning behind the suggestion and from which area of expertise it stemmed. This insight into human-human collaboration suggests that it is by no means a requirement that a computational tool only returns output that is aligned with the users' perspective, expectation, or established epistemologies. For such misalignments to be productive, however, the user needs to be enabled to engage with the tool's output purposefully and intentionally. We observed that participants tended to expect high levels of accuracy from the CV-based tool, which sometimes materialized in over-reliance on the tool's output. We suggest addressing this known effect in human-AI collaboration (see, e.g.,~\cite{Buccinca_Trust_Think2021}) in such a way that encourages users to scrutinize the output of a tool as a `suggestion' that they can contradict and potentially refuse. 

Being able to understand how computational tools function is considered a prerequisite for their `meaningful' integration into humanities research~\cite{koolen_toolcriticism_2018}. 
However, we should not shift the responsibility to acquire the needed skills to the users of computational tools. Our findings confirm that in order to enable `meaningful' human-AI collaboration, we need to enable critical reflection \emph{during} the users' interaction with a tool (see Section~\ref{crtoolsupport}). Based on these findings, we suggest approaching the design of human-AI collaboration with a strong emphasis on `supporting reflection on the basis of transparency.' Transparency in this context refers not only to the clear presentation of the functional scope of a tool (see, e.g.,~\cite{Amershi_2019_GuidelinesAI}) but also extends to making it obvious for the users that their active scrutinization and critical reflection constitute a central and significant element of interaction. As our findings show, puzzlement about the tool's results caused some participants to engage in hermeneutic interpretation when attempting to derive meaning from the tool's results, which is in line with the suggested effects of microboundaries, i.e. small obstacles~\cite{Coxetal2016}, and breakdowns as intentional violations of the users' expectation~\cite{baumer2015reflective}. However, we also observed that perplexity could negatively influence the participants' willingness to engage in critical reflection and interpretation. Hence, we suggest supporting reflection by transparently framing the interaction with a computational tool as something that will include instances of perplexity while also encouraging users to tap into the productivity of their own algorithmic imaginary.

\subsubsection{Foregrounding epistemic presumptions}

Assumptions about a computational tool are formed prior to interacting with it and are also influenced by public perceptions and assertions about technology, for example, about the capabilities of AI-based systems. As mentioned above, our study emphasizes the importance of clearly communicating a tool’s functional scope and limitations, which resonates with Amershi et al.’s general guidelines for designing human-AI interaction~\cite{Amershi_2019_GuidelinesAI}. However, this would not be sufficient in research application contexts since it does not address the scope and limitations of a tool in regard to underlying epistemic presumptions. Art history, just like other disciplines, incorporates a variety of different research approaches, methods, epistemic presumptions, and associated positional differences. However, our study participants tended to foreground the thematic scope of their research and practice without expanding on the epistemic presumptions and methodological approaches this entails and refers to. 
Even if a clearly communicated epistemic self-positioning did not occur, it still shaped and influenced their interactions with the CV tool. This implies that critical-reflective human-AI collaboration also requires a component of self-awareness or self-reflection (see~\cite{Mortari_2015}). Self-reflection, in this case, would be aimed at reducing a misalignment of epistemology, which could have a negative effect on the perceived usefulness of a computational tool and inhibit `meaningful' human-AI collaboration.

Hence, we derive the implication that computational tools should not only adhere to the principles of transparency laid out above but also encourage the users to make explicit their own epistemic presumptions (see~\cite{DIgnazio_Klein_2020, agre1997criticaltechnicalpractice}). As our findings confirm, humanities researchers and art historians value the potential of computational tools to influence epistemology and knowledge structures (see also~\autoref{sec:CV_ArtHistory}). However, this effect creates friction during usage when it coincides with a lack of awareness about the conflicting epistemologies. Consequently, critical-reflective human-AI collaboration hinges on the users' awareness regarding the purpose of their tool usage and the epistemic presumptions with which they approach a tool. We envision that this design implication could, for example, be operationalized by extending the tool with an ``onboarding'' interaction sequence that would guide the user through self-reflection regarding their epistemic presumptions and intended purpose of their interaction with a tool. This self-positioning would be mirrored with relevant information about the tool that also makes transparent that tool output might not match the users' initial expectation, i.e., explicitly encouraging critical reflection and scrutinization (see also Section~\ref{sec:DItransparency}). 

\subsubsection{Emphasizing the situatedness of data}

When applying computational tools to highly selected and canonized data, as is the case in art history, it is essential to emphasize the historicity and situatedness of the data and information and enable a critical-reflective perspective on both. D’Ignazio and Klein~\cite{DIgnazio_Klein_2020} highlight that the mechanisms for the production of data - which include social, cultural, historical, and material conditions - need to be disclosed (see~\autoref{sec:CV_ArtHistory}). 
Our study also echoes this claim. As one participant put it:
\begin{displayquote}
\textit{``How were the data collected, and what were the exclusion criteria? Was it even reflected that possibly only one-sided data could be fed in? So, what was available, and what was in the public domain? These are all aspects for which I would like to see appropriate information as a disclaimer preceding the use.''}
(P5)
\end{displayquote}
Several of our interviewees associated `objectiveness,' `completeness,' and `reliability' with data-based and computational approaches. This association also influences the perception of a tool. In order to counteract this, we suggest placing a much bigger focus on making explicit what kind of data, i.e., which collections, can be accessed through a tool. 
Purposefully selecting a collection that matches a given search interest presupposes that the users have contextual knowledge about a collection's implicit and explicit biases, i.e., the focus, historical development, and canon ingrained in it. However, this contextual knowledge might not exist when the user is confronted with an unknown collection or an aggregated dataset. Emphasizing the situatedness of data could build on approaches such as Gebru et al.'s suggestion to include ``Datasheets for Datasets''~\cite{Gebru_Datasheets2021} or take inspiration from Holland et al.'s ``Dataset Nutrition Label''~\cite{Holland_etal_2018}. Instead of having the users start with a pre-selected collection as default, the selection could be designed as a required interaction that would be enabled based on the included contextual information about the underlying data.

\subsubsection{Strengthening interpretability through contextualized explanations}

The tool's lack of interpretability prevented participants from fully realizing their aspiration to reflect on methods: they could not reflect on how the tool's output came about and how the output relates to their search. The question that arises from this is, first, how we can enable interpretability of a CV tool - considering that we envision stakeholders with little to no technical understanding of CV. Second, we need to factor in that domain experts interpret AI technologies within and against a diverse backdrop of existing themes and practices~\cite{Benjamin_ExplStrategies2022}. Current research in the area of explainable AI or human-centered explainable AI (HCXAI) (see ~\cite{EhsanUpol2022}) addresses the question of how AI-based applications can be made interpretable, which is particularly challenging when considering `nontechnical' users. As pointed out by Ehsan et al.~\cite{EhsanUpol2022}, HCXAI needs to consider the explanation needs of the user, the purpose of explanations, and where and when the explanations are provided. As Benjamin et al. have shown, stakeholders with no expertise in AI approach explanations with their own situated sensemaking~\cite{Benjamin_ExplStrategies2022}. Benjamin et al. suggest, amongst other things, offering ``contextual cues'' that enable stakeholders to combine or contrast a given explanation with their own explanation strategies~\cite{Benjamin_ExplStrategies2022}. 

Based on this related work and our findings, we imply that critical-reflective human-AI collaboration requires `contextualized explanations.' Such explanations would consider domain-specific practices of interpretation and sensemaking instead of purely focusing on a technological understanding of algorithmic explanation methods. As discussed above, some participants applied a hermeneutic approach to interpreting the tool's output: they tested and observed the effect of different embeddings, i.e., different `concepts' of similarity encoded in the algorithms, and tried to deduce what visual features or visual information of the input image were picked up in the tool's output. Our suggestion of `contextualized explanations' aims to support and encourage such an interpretative and `hermeneutic' approach. 
We suggest addressing interpretability by designing contextualized explanations that relate the effects of algorithms, machine learning models, and data on a tool's output to the user's situated sensemaking. Contextualized explanations will require a thoughtful mapping of concepts into the stakeholders' vocabulary. 


\section{Limitations}
Due to the specific focus of our research and our sample size, we do not claim that our results can be generalized to other, completely different contexts. However, since critical reflection constitutes a core aspect of all scientific and scholarly work, we do assume that our findings can inform the design of human-AI collaboration in other contexts and disciplines as well. Our findings' core aspects align with existing research into human-AI collaboration in qualitative contexts (e.g.,~\cite{Jiang_serendipity, Feuston_Tools, chen_ambiguity, Amershi_2019_GuidelinesAI}), showing us that there is a reasonable overlap despite the differences in contexts. 

We are aware of the limitations that relate to potential biases arising from our study participant selection. Our study relied on voluntary participation, which might induce bias insofar as potential participants would only agree to be interviewed if they had at least some degree of interest and openness toward the topic of human-AI collaboration. However, of the initial 17 potential participants we contacted, only five declined participation, each with reference to their current workload. 
At the same time, compensating participants could likewise engender positive feelings that collude feedback~\cite{Pater_ParticipantCompensationHCI}. 
Additionally, in technology-based studies, ``novelty bias'' could influence the study outcome when participants are excited to try something new~\cite{Ming_ResponseBias}. Hence, we made sure that our sample included participants with different levels of familiarity with computational approaches (see \autoref{sec:methods} and \autoref{tab:participants}). 

As pointed out in \autoref{sec:UseCaseComputerVision}, there exist only a few implementations of ready-to-use CV tools specifically intended for nontechnical users that give access to art historically relevant image collections. Accordingly, development is still in an exploratory and prototypical phase, which also means that existing tools and interfaces might not be as well-refined as other long-term infrastructural information retrieval systems, for example, in terms of usability. This is also the case for the tool we used in the software exploration, which we acknowledge as a limitation of our study. 

Despite these limitations, our work opens promising directions for future research.

\section{Conclusions and Future Work}

With the work presented in this paper, we suggest the realization of `meaningful' human-AI collaboration by integrating it with the concept of `designing for reflection.' We conceive criticality and reflection as genuinely human competencies that are central to scholarly knowledge production. We explore the question of how critical reflection can be enabled in human-AI collaboration with a use case situated in humanities research. Specifically, our use case focuses on computer vision (CV)-based tools for art historical image retrieval. We conducted a qualitative interview study and think-aloud software exploration with 12 art historians. Our findings confirm our hypothesis that critical reflection needs to be actively supported and scaffolded during interaction with a computational tool. Based on the insights derived from our study, we suggest four design implications for enabling critical reflective human-AI collaboration: supporting reflection on the basis of transparency, foregrounding epistemic presumptions, emphasizing the situatedness of data, and strengthening interpretability through contextualized explanations. 

Technological advancements in AI are increasingly becoming available to `nontechnical' users and domain experts through ready-to-use tools, which also affect and mediate their understanding of the world. Verbeek emphasizes that ``[u]sers, designers, and policymakers should be enabled to read, design, and implement technological mediations, in order to be able [to] deal in a critical, creative, and productive way with powers that remain hidden otherwise"~\cite[p. 31]{Verbeek_Mediation2015}. We contribute to such efforts by highlighting that computational tools, i.e., AI, challenge critical reflection. Acknowledging this challenge is not only important in the humanities but can be extended to other contexts. Awareness of how the use of computational tools affects, shapes, and changes epistemic assumptions and research processes is necessary across all forms of knowledge production. Hence, our work has relevance for current discourses concerned with the question of how we can integrate AI in research and education in a `meaningful' way. Such discourses are informed by work on ethical and responsible realizations of AI. In this regard, transparency and interpretability are being discussed as core principles that need to be accounted for. Our empirically grounded design implications mirror the importance of these aspects. However, just like tool development, the challenge of enabling transparency and interpretability tends to be addressed from an engineering perspective that prioritizes metrics like usability and efficiency. In contrast, our research suggests approaching tool design in a highly contextualized manner that builds on a nuanced understanding of what constitutes `meaningful' human-AI collaboration in the designated area of application. The successful implementation of such an approach depends on a close interweaving of theory, empirics, and technology development. Hence, we do not claim that our results are generalizable, but we do believe that our research is a valuable contribution to the question of how we can approach the realization of `meaningful', i.e., critical-reflective human-AI collaboration, especially in scholarly and educational contexts.  

Our future work aims at operationalizing our design implications. We are currently conducting a follow-up study to translate our findings into concrete ``explanation needs.'' This future work relates to one of our design implications: We will probe how we can bridge the users' explanation needs with the technical capabilities of explanation methods for computer vision in order to provide ``contextualized explanations''. We will further explore this topic through participatory design workshops with domain experts from the humanities and HCI.


\bibliographystyle{ACM-Reference-Format}
\bibliography{bibliography}

\appendix
\section{Appendix: Interview Protocol}
\label{InterviewProtocol}
[\emph{The following protocol is a translation of the entirety of the script (originally in German). It was slightly adapted to ensure anonymity of the authors.}]

\subsection*{Welcome and contextualization}

Thank you for participating in this study! [At the HCI research group], we conduct research on the design of systems that contain so-called ``artificial intelligence''. We are interested to conceptualize the design in such a way that collaboration between humans and ``AI''-based systems is enabled. For this purpose, it is important that we understand the demands and needs of (future) users. One of our current use cases [$\ldots$] deals with the design of ``AI''-systems for humanities research. More precisely, we investigate the design space for tools that can be used in the context of art history. We conduct this interview because we are interested in your expertise as an art historian. We want to, first of all, get a better picture of your research and work practice. In the `software exploration', which will take place in the second part of our interview, we want to gather your impressions and thoughts about an existing web-based tool that integrates computer vision (we do not develop this tool, but use it as an example in our interview). For the software exploration, I had invited you to bring your own example image - have you already had the opportunity to select an image that is relevant to your research or in the context of your current work? It would be great if you have it readily available on your computer, for example on your desktop. You will use it later during the software exploration. 

\subsection*{Consent}
You have already given your written consent to participate in this study and also agreed that I may record our interview. I will use the ``record'' function of Webex for this purpose. The recordings will be stored locally on my university computer and be transferred to a password-protected university-owned server. Only I and my colleagues will have access to the recording for the analysis of the interviews. Would you be so kind to confirm your consent? 

\subsection*{Interview Script}
\begin{enumerate}
   \item To begin, I would like to ask you to briefly describe the field in which you work as an art historian.
   \item What is your current thematic focus? For example, what is your current research topic or work interest?
   \item When you think about a typical research project or another kind of art historical [content-oriented] work activity of yours: what role does collaborating with other people play in this?
        \begin{itemize}
            \item Can you be more specific about the concrete research tasks or content-oriented work activities for which you collaborate with others?
            \item(\emph{If answer remains on a meta-level or only refers to administrative/managerial tasks}) I would like to encourage you to think more in the direction of research-related or scholarly tasks and content-oriented work activities: How is the collaboration structured when you engage in work that is ``art historical'' in nature? Could you, for example, describe a concrete workflow?
        \end{itemize}
   \item What role do computational tools or computational [research] approaches (for example approaches that build on ``AI'') play for your research and content-oriented work? 
        \begin{itemize}
            \item (\emph{If unspecific/unclear}) Which computational tools/approaches do you use regularly in your research or for content-oriented work? For which specific tasks do you use them?
                \begin{itemize}
                    \item[$\circ$] (\emph{If none}) Can you tell me your reasons for not using computational tools?
                \end{itemize}
        \end{itemize}
    \item Image-based search engines or other applications that integrate so-called ``computer vision'' or ``machine vision'' are currently being explored in the context of art history. What is your experience with this technology so far?
        \begin{itemize}
            \item (\emph{If only ``commercial'' search engines like Google (reverse) image search are mentioned}) Do you also know image-based search engines that are specifically deployed for art historical research contexts?
        \end{itemize}
    \item Apart from your own experience with this technology: how do you assess the use of ``computer vision'' or other ``AI''-based applications in art historical contexts in general?
        \begin{itemize}
            \item (\emph{If too unspecific}) How do you assess, for example, the potentials and/or challenges associated with computer vision and/or ``AI''? Could you expand on this and deepen what you have just said in this respect?
        \end{itemize}
\item You have already (implicitly) talked about the (general) potentials and challenges that you associate with computational tools and approaches - what aspects are particularly important to you from your perspective as an art historian?
    \begin{itemize}
        \item (\emph{If no reference to specific competencies is made}) Which competencies are most important for you in connection with the use of computational tools? Put differently: from your point of view, which competencies should each and every art historian bring to the table when using computational tools in their research and art historical work?
    \end{itemize}
\item When you think again about the computational tools or approaches that you have talked about so far: What do these tools/approaches not yet accomplish, but which would be important for your research or art historical work?
\item Do you wish to add anything before we proceed with the software exploration?
\end{enumerate}

\subsection*{Software Exploration}

\subsubsection*{Familiarization and exploration of the tool}
The goal of the software exploration is to better understand how you go about using a computational tool. In particular, we are interested to hear your thoughts, opinions and impressions while using the tool \texttt{imgs.ai} - this is a web-based search tool that integrates computer vision. It was developed by a group of researchers at FAU Nürnberg [participants are again reminded that this tool is not being developed by the authors]. The tool allows you to conduct image-based searches in collections. I'll send you the link to the tool in the chat now. I would ask you to share your screen with me so that I can follow you during your interaction with the tool. Please also ``think-aloud'' while using the tool and simply describe your spontaneous impressions and associations or what questions or thoughts are going through your mind. It's best to imagine that I'm not seeing what you're doing at all and have to rely on your verbalization of what you're doing or what's going through your mind. Please keep in mind that I do not evaluate you or how ``well'' you interact with the tool - I am interested to hear more about your impressions, thoughts and opinions that come up while you use the tool. 

[Let them explore the functionalities of the tool for about 3-4 minutes - advise/help in case participant is not able to operate tool without further instruction.]

\subsubsection*{Image retrieval task}

\begin{itemize}
    \item (\emph{If participants do not automatically proceed with their image retrieval task, encourage them to use ``upload'' function of the tool}) 
    Would you care to explain to me what your search interest is?
    \begin{itemize}
        \item [$\circ$] Why have you chosen this image?
        \item [$\circ$] What kind of images do you wish/expect to retrieve?
        \item [$\circ$] What would be a ``relevant'' result?
        \item [$\circ$] To what research question or content-oriented work does this image/your search relate to?
    \end{itemize}
\end{itemize}  

[Let them refine search results for about 3-4 min]

End of the exploration

\subsection*{Outro questions}
\begin{enumerate}
    \item Can you imagine - in principle - to integrate such a tool into your research / use such a tool in your art historical work?
        \begin{itemize}
            \item [$\circ$] If yes: for which tasks?
            \item [$\circ$] How would you go about it?
        \end{itemize}
    \item Thank you very much....We have now reached the end of the interview. Is there anything else we haven't touched on so far that you would like to mention?
\end{enumerate}

\end{document}